\documentclass[sigconf,screen,10pt]{acmart}

\AtBeginDocument{%
  }
\begin{document}
\setcopyright{none}
\settopmatter{printacmref=true, printccs=false, printfolios=true}
\renewcommand\footnotetextcopyrightpermission[1]{}
\acmDOI{}
\acmISBN{}
\acmArticle{}
\acmPrice{}
\acmConference{Preprint}{arXiv Submission}

\title{ASC-Hook: fast and transparent system call hook for Arm}

\author{Yang Shen}
\affiliation{
  \institution{National University of Defense Technology}
  \city{Changsha}
  \state{Hunan}
  \country{China}}
\email{shenyang_23@nudt.edu.cn}

\author{Min Xie}
\affiliation{
  \institution{National University of Defense Technology}
  \city{Changsha}
  \state{Hunan}
  \country{China}}
\email{xiemin@nudt.edu.cn}

\author{Tao Wu}
\affiliation{
  \institution{Changsha University of Science and Technology}
  \city{Changsha}
  \state{Hunan}
  \country{China}}
\email{wutao@stu.csust.edu.cn}

\author{Wenzhe Zhang}
\authornote{Corresponding author.}
\affiliation{
  \institution{National University of Defense Technology}
  \city{Changsha}
  \state{Hunan}
  \country{China}}
\email{zhangwenzhe@nudt.edu.cn}
\begin{abstract}

Intercepting system calls is crucial for tools that aim to modify or monitor application behavior. However, existing system call interception tools on the ARM platform still suffer from limitations in terms of performance and completeness. This paper presents an efficient and comprehensive binary rewriting framework, ASC-Hook, specifically designed for intercepting system calls on the ARM platform. ASC-Hook addresses two key challenges on the ARM architecture: the misalignment of the target address caused by directly replacing the SVC instruction with br x8, and the return to the original control flow after system call interception. This is achieved through a hybrid replacement strategy and our specially designed trampoline mechanism. By implementing multiple completeness strategies specifically for system calls, we ensured comprehensive and thorough interception. Experimental results show that ASC-Hook reduces overhead to at least 1/29 of that of existing system call interception tools. We conducted extensive performance evaluations of ASC-Hook, and the average performance loss for  system call-intensive applications is 3.7\% .
\end{abstract}


\maketitle

\section{Introduction}
System calls are a crucial mechanism for interaction between user programs and the operating system kernel, typically used to request higher privileges from the operating system to perform critical tasks. Intercepting system calls has various applications, such as (i) tracing and debugging, (ii) enhancing program reliability and security, (iii) simulating different operating system environments, and (iv) adding binary compatibility to support new operating system subsystems. 

In this context of system interaction and computing efficiency, the ARM architecture is widely used in various embedded devices, while ARM64, with its high performance and energy efficiency, is extensively applied in fields such as the Internet of Things (IoT), edge computing, autonomous driving, and communication infrastructure.

Historically, there has been significant research focused on addressing the problem of system call interception. One category of this work leverages the fact that system calls have a unified execution path within the kernel. A typical example is Ptrace\cite{padala2002ptrace}, which can extensively intercept system calls. However, this method requires frequent invocations to manage process execution or monitor register states, leading to frequent kernel transitions and significant performance loss.
Binary rewriting is a method of modifying the semantics of a program by directly rewriting the compiled binary code without source code.Typical binary rewriting tools, such as Pin\cite{10.1145/1065010.1065034} and Multiverse\cite{bauman2018superset}, offer comprehensive syscall interception capabilities.Multiverse maps original instruction addresses to modified instruction addresses through a two-level address translation table, but this data structure still incurs significant overhead. Pin inserts binary code for instrumentation at runtime but requires binary analysis and code translation, resulting in significant performance overhead. Another method is to replace the system call instructions with debug instructions or illegal instructions, thus intercepting through signals. System Call User Dispatch (SUD) \cite{bertazi2021syscall}is a new mechanism introduced in Linux that essentially intercepts system calls through the SIGSYS signal. Although these methods using signal interception are comprehensive, they cause a certain amount of performance overhead due to the frequent kernel transitions required for signal handling.
\begin{figure}[!t]
  \centering
  \includegraphics[width=0.5\textwidth]{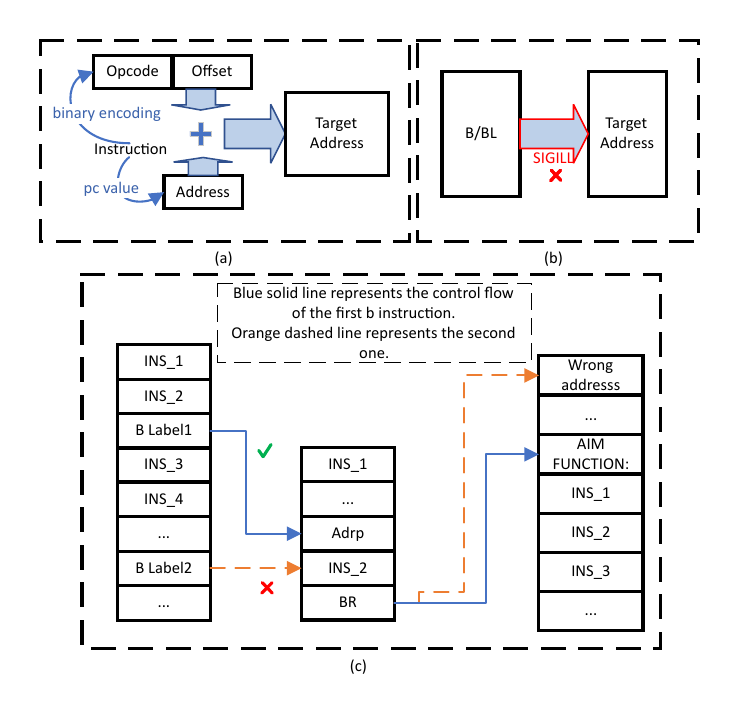}
  \caption{(a) In PC-relative addressing mode, the effective address is obtained by adding the offset to the PC. (b) The offset between the PC value of the b or bl instruction and the target address is beyond the encodable range. (c) ADRP is the first replaced instruction, and it is used to load the target jump address into the destination register. BR is the second replaced instruction. The first b instruction successfully reaches the jump target by executing both the ADRP and BR instructions. However, the second b instruction, which falls between the ADRP and BR instructions, skips the execution of the ADRP instruction, resulting in a jump to the incorrect target address.}
  \Description{A brief description of the image.}
  \label{fig:combined}
\end{figure}
The aforementioned methods, to varying degrees, struggle to balance both completeness and efficiency in system call interception. Recently, Kenichi Yasukata and Hajime Tazaki proposed a new method for system call interception on the x86 architecture—zpoline\cite{288689}. Since the rax register in the x86 architecture holds the system call number, they achieved system call interception by replacing the syscall/sysenter instructions with the two-byte callq *\%rax instruction. This method successfully balances efficiency and completeness.

However, This research\cite{288689} primarily focuses on the x86 architecture and also points out that this approach faces the following two main challenges on the ARM instruction set:
\begin{itemize}
\item 
"zpoline is not compatible with CPU architectures which assume the instructions to be aligned by

architecture-specific sizes on the memory and consider a jump to an unaligned virtual address as an invalid operation (e.g., ARM); this is because, when zpoline is applied, the execution can jump to an unaligned virtual address between 0 and the maximum system call number."\cite{288689}
\item 
"Besides the issue of the instruction alignment, binary rewriting techniques need to pay attention to architecture-specific factors; for example, on ARM CPUs, the simple replacement from SVC to BL overwrites/breaks the return address saved in a specific register."\cite{288689}
\end{itemize}

To address the first challenge and ensure the jump distance, we propose a solution that combines multiple Replacement strategy. To solve the second challenge, we design a dedicated trampoline entry for each SVC instruction to save the return address. Leveraging the nature of system calls, we developed a specialized mechanism to ensure the completeness of system call interception, using multiple completeness strategies to guarantee the security and thoroughness of the algorithm. Our method not only solves the issues of address alignment and jump distance but also avoids disrupting critical registers, preserving the return address. This enables efficient and comprehensive system call interception on the ARM architecture.

Specifically, this paper makes the following contributions:

\begin{itemize}
\item 
We address two key challenges previously mentioned in system call interception on the ARM architecture: the issue of misaligned jump target addresses and the problem of returning to the original control flow after interception.
\item 
We observe two key properties: for the vast majority of system calls, at least two instructions can be replaced, and the total number of SVC instructions in a process's image is very limited. We substantiate these observations with extensive statistical data.
\item 
We implement an interception scheme on the ARM architecture that replaces two instructions, achieving efficient system call interception while minimizing memory consumption.
\item 
For extreme cases, we designed a dedicated mechanism for system calls that ensures the completeness of the interception algorithm.
\end{itemize}

To the best of our knowledge, this work is the first to efficiently achieve ARM system call interception with completeness, efficiency, transparency, and without requiring source code. We extensively tested ASC-Hook, evaluating it with several commonly used system call-intensive applications, including Redis\cite{sanfilippo2009redis}, Apache\cite{apache}, IOR\cite{llnl2015ior} \cite{ior}, Nginx\cite{nginx}, and SQLite\cite{sqlite}. The experimental results demonstrate that ASC-Hook provides a significant performance improvement compared to common system call interception methods such as ptrace and signal interception. In extreme cases, the overhead of ASC-Hook is one-sixtieth of ptrace. Based on our extensive testing, ASC-Hook maintains an average performance overhead of 3.7\% for most system call-intensive applications.

We plan to open-source our tool and provide documentation for the community upon acceptance of the paper, subject to the completion of the review process.

\begin{figure}[!t]
  \centering
  \includegraphics[width=0.50\textwidth]{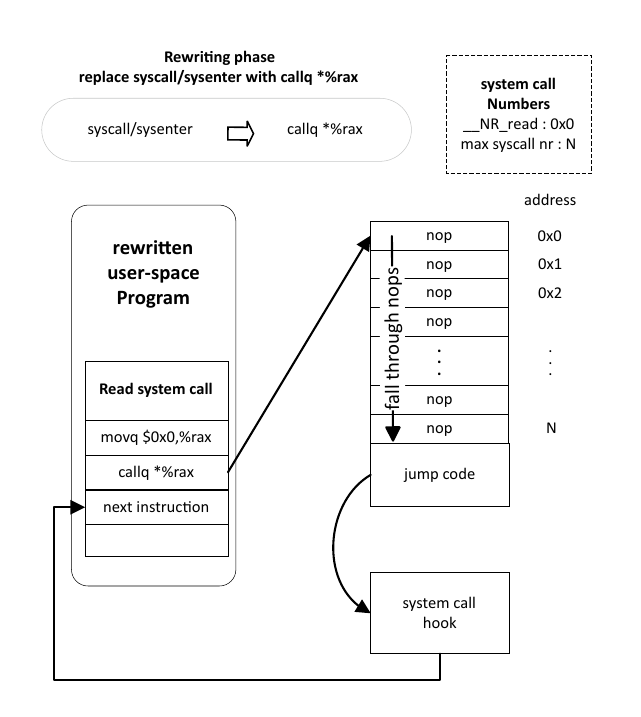}
  \caption{ 
An example execution flow of zpoline for the read system call}
  \Description{A brief description of the image.}
  \label{fig:figure9}
\end{figure}

\section{Motivation}

Recently, Kenichi Yasukata and Hajime Tazaki proposed a new method for system call interception on the x86 
architecture
——zpoline\cite{288689}. As shown in Figure \ref{fig:figure9},the rax register holds the system call number before the system call, and by replacing the syscall/sysenter instruction on the x86 architecture with the two-byte callq *\%rax instruction,it allows jumping to virtual addresses between 0 and the maximum system call number. By instantiating trampoline code at virtual address 0, this method allows for efficient system call interception by modifying one instruction to jump to the trampoline function.Notably, zpoline is the leading tool on the x86 platform, offering unmatched efficiency and completeness in system call interception.

However, this approach faces challenges on the ARM platform. The first challenge is that the ARM architecture requires instruction addresses to be 4-byte aligned. If a similar instruction to callq *\%rax on x86 is directly used, it could lead to a jump to a non-aligned address since system call numbers are not always multiples of 4, causing a bus error. Additionally,as shown in Figure \ref{fig:combined}(b), directly replacing the svc instruction with a b instruction often faces the challenge of insufficient jump distance.

The second challenge concerns how to return after the jump. The ARM architecture does not have an instruction like callq that can jump and push the return address onto the stack. Using the BL instruction would overwrite the important x30 register, which is clearly unacceptable.

This paper is based on two key observations. 
\begin{itemize}
\item The first observation is that we have noticed that for most system calls, at least two instructions can be replaced. This is because, on the ARM architecture, the system call number must be passed to the x8 register before executing the SVC instruction. If the system call has additional parameters, registers x0 to x7 need to be assigned with system call parameters or passed to the kernel through stack operations. Therefore, we can at least replace the svc instruction and the instruction that assigns the value to the x8 register.

\item Our second observation is that the total number of SVC instructions within a process's memory image is very limited. This is especially evident for most high-level language users who typically do not write SVC instructions directly but instead use system calls provided by existing shared libraries. Furthermore, the number of system calls in the Linux operating system is currently fewer than 600. Table \ref{tab:table1} shows the number of svc instructions in the process images of some of the applications or commands we tested, with most having fewer than one thousand\footnote{The testing environment for the applications in Table \ref{tab:table1} and Table  \ref{tab:svc_instructions_compact} is consistent with the environment described in the evaluation section of this paper. The evaluation section also includes the testing methods for most of the applications listed in the tables. For more details, please refer to the evaluation section.}. The number of SVC instructions in the process images of these applications is roughly the same because they use similar dynamic libraries (such as glibc)\footnote{The primary sources of the SVC instructions listed in Table \ref{tab:table1} are the dynamic linker (ld-linux-aarch64.so.1), the main glibc library (libc-2.31.so), and the POSIX thread library (libpthread.so.0). These commands contain fewer SVC instructions compared to other applications because they do not rely on the POSIX thread library.}.
\end{itemize}
Based on the first observation, we addressed the first challenge. Our method replaces the SVC instruction and the instruction preceding the SVC that assigns a value to the x8 register. We address this challenge by assigning x8 an immediate value aligned to 4 bytes and then executing the br x8 jump instruction. Detailed descriptions are provided in Section \ref{sec:section3.1}.

\begin{table}[h!]
\centering
\begin{tabular}{|l|c|}
\hline
\textbf{Command/Program} & \textbf{The number of svc } \\
\hline
cat command & 727 \\
\hline
touch command & 727 \\

\hline
bfs(mpi) & 929 \\
\hline
Redis\cite{sanfilippo2009redis} & 929 \\
\hline
nginx\cite{nginx} & 929 \\
\hline
IOR\cite{llnl2015ior} \cite{ior} & 929 \\
\hline
Apache\cite{apache} & 929 \\
\hline
Hacc-io\cite{hacc-io} & 929 \\
\hline
SQLite\cite{sqlite} & 929 \\
\hline
\end{tabular}
\caption{The number of svc instructions in the process images}
\label{tab:table1}
\end{table}

Based on the second observation, we addressed the Second challenge. We custom-designed a dedicated trampoline for each SVC instruction, with each trampoline corresponding one-to-one with each SVC instruction in the process image. In this trampoline, we save the return address information, which is the address of the next instruction following the SVC instruction, thereby ensuring the correct return path. Additionally, through our multi-level trampoline design, we minimize the code length required for each SVC instruction. Given that the number of SVC instructions in a process image of typical applications is relatively small, our trampolines do not introduce significant memory and virtual memory overhead, as detailed in Section \ref{sec:section3.2}.

However, because our approach requires replacing two instructions (the SVC instruction and the instruction preceding it that assigns a value to x8), we noticed that in rare cases, some libraries contain only the SVC instruction without an x8 assignment instruction within the preceding 20 instructions. In the applications we tested, among the SVC instructions in the process image, two SVC instructions exhibited this situation. Additionally, since our approach requires replacing two instructions, it is crucial that there is no jump target address between the replaced instructions(including the second replaced instruction). As shown in Figure  \ref{fig:combined}(c), ADRP and BR are the replaced instructions. The ADRP instruction is responsible for loading the jump target address into the BR instruction's destination register. If the jump target address appears between the ADRP and BR instructions (including the BR instruction, excluding the ADRP instruction), the target address will not be loaded into the destination register, leading to an incorrect jump address. Due to the difficulty in determining the target address of indirect jump instructions, this problem has not been fully resolved across different architectures \cite{10.1145/3385412.3385972}\cite{lei2023birfia}.

We first perform static analysis of the instructions preceding the SVC and the jump instructions within the process image to identify the above two situations. For the issue of indirect jump target addresses, which cannot be resolved through static analysis, we propose a novel analysis strategy specific to system calls. Specifically, in our replacement strategy, we only replace two instructions, not three. This applies even for system calls with parameters. In such cases, we only need to consider the scenario where the first instruction is not executed, meaning that only the BR instruction is executed. When only the br x8 instruction is executed, since x8 contains the system call number, this jump will inevitably cause a segmentation fault or bus error. We then perform further analysis within the signal handler for segmentation faults or bus errors to confirm the problematic SVC instruction. For all identified cases, we replace these SVC instructions with brk or illegal instructions to intercept them using signals, as detailed in Section \ref{sec:section3.3}.
\begin{table}[h!]
\centering
\begin{tabular}{|l|c|c|}
\hline
\textbf{Application} & \textbf{Used svc } & \textbf{svc requiring signal  } \\

\hline
bfs(mpi)     & 75 \& 97  & 0 \\
\hline
Redis   & 60        & 0 \\
\hline
nginx   & 61        & 1 \\
\hline
IOR     & 81 \& 97  & 0 \\
\hline
Apache  & 60        & 1 \\
\hline
Hacc-io & 82 \& 97  & 0 \\
\hline
SQLite  & 17        & 0 \\
\hline
\end{tabular}
\caption{Number of svc instructions used by applications and those requiring signal interception}
\label{tab:svc_instructions_compact}
\end{table}

We conducted tests on the applications listed in Table \ref{tab:svc_instructions_compact}. The second column of the table represents the actual number of svc instructions required by the application in the process image. We distinguish these svc instructions based on their different PC values. It is worth noting that the applications create multiple processes or threads, and most processes or threads use roughly the same number of svc instructions. However, BFS, IOR, and HACC-IO utilize the exec system call, which results in significant differences in the number of svc instructions used between certain processes or threads. Therefore, for these three applications, we present multiple values to reflect the varying numbers of svc instructions required by different processes. \footnote{Since we may not have traversed all branches of the applications, it can be assumed that the number of svc instructions required by the application is at least as many as shown in the table.}
The third column shows the number of svc instructions that must be intercepted using the signal handling method. We observe that only Nginx and Apache have such an svc instruction. This further validates our first observation that, for the vast majority of system calls, at least two instructions can be replaced. As for the single svc instruction in Nginx and Apache, it can be detected using our first completeness strategy, which is discussed in detail in Section \ref{sec:section3.3}. Additionally, we examined all of the 929 svc instructions in the process image and found 7 instances where the jump target address is located between two replaced instructions. However, these cases can also be detected using our completeness strategy.

\begin{figure*}[!t]
  \centering
  \includegraphics[width=\textwidth,height=0.4\textheight,keepaspectratio]{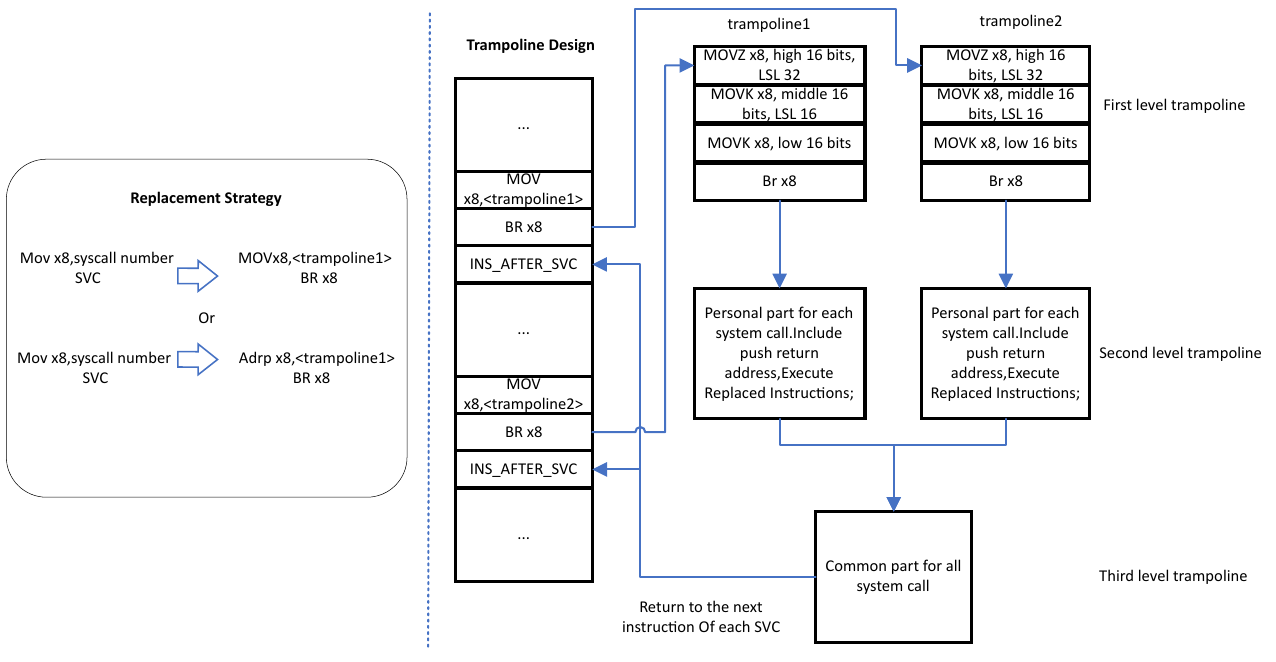}
  \caption{Replacement Strategy and the execution process of each system call after interception}
  \Description{A brief description of the image.}
  \label{fig:figure2}
\end{figure*}

\section{Methodology and Implementation}
\subsection{Replacement Strategy}
\label{sec:section3.1}
Our replacement strategy includes three methods:
\begin{itemize}
\item  Using the mov instruction to fill the target address into the destination register x8 and replacing the SVC instruction with a br instruction. This method is applied to the first 3,840 SVC instructions found in the process image and is currently our primary replacement method.

\item Using the adrp instruction to fill the target address into the destination register x8 and replacing the SVC instruction with a br instruction. When the number of SVC instructions in the process image exceeds 3,840, this method is used for replacing the portion beyond 3,840. Although we have not yet encountered such applications, this serves as a precautionary measure.

\item   Directly replacing the SVC instruction with an illegal instruction or brk instruction. This method is applied in some extreme cases, which will be specifically explained in Section \ref{sec:section3.3}.
\end{itemize}

We first introduce the first two replacement methods. As mentioned above, due to the special ABI of system calls, we have at least two instructions available for replacement: one is the SVC instruction itself, and the other is the inevitable assignment to the x8 register.

Firstly, we choose to use x8 as the register to store the jump target address. This is because, according to the system call ABI, there must be an instruction that assigns the system call number to x8 before the SVC instruction. This means we can find this instruction before the SVC instruction and re-execute it in the subsequent trampoline function to restore the value of x8 register. The choice of x8 as the target register is also related to our completeness strategy, which will be specifically explained in Section \ref{sec:section3.3}. We search backward from the SVC instruction to find the assignment statement to the x8 register and use this instruction as one of our replacement instructions.

We replace the SVC instruction and the previously found assignment instruction to the x8 register. The preceding instruction is replaced with a mov instruction to x8, with the immediate value being the address of the first-level trampoline, and the SVC instruction is replaced with br x8. This method has certain limitations because ARM architecture instructions are typically 4 bytes and the immediate value for the mov instruction is only 16 bits, meaning the jumpable virtual address range is from 0 to 65535, accommodating only 16383 instructions. According to our trampoline design, this can accommodate up to 3840 trampolines. As we previously observed, the total number of SVC instructions is very limited, making it almost impossible to reach the 3840 trampoline limit. The design of the trampoline will be explained in detail in Section \ref{sec:section3.2}.

The second method replaces the assignment instruction to x8 with an adrp instruction and the SVC instruction with br x8. The adrp instruction takes a 21-bit signed immediate value, shifts it left by 12 bits to form a 33-bit signed number, clears the lower 12 bits of the PC address to get the current page address, and adds them together to compute the target page address, which is then written to a general-purpose register. Therefore, after the adrp instruction, the address in x8 is 4KB aligned. If we use this method, we must place the trampoline at the start of a page, which results in significant memory waste. Thus, we use the second method only when the first method reaches its 3840 trampoline limit.

Through the above replacement methods, we solve the issue of the b instruction's insufficient relative jump distance and the problem of misaligned jump addresses in the zpoline\cite{288689} method when directly replacing with br x8 on the ARM architecture.

\subsection{Trampoline Design}
\label{sec:section3.2}
Our trampoline design aims to solve two critical issues. The first issue is how to return to the original instruction flow. This is specific to the ARM architecture because the x86 callq instruction pushes the return address onto the stack, whereas the ARM blr instruction stores the return address in the x30 register. If we directly replace the SVC instruction with a blr instruction, the value in the x30 register, which contains the function's return address, would be lost. Therefore, we must use the br instruction to jump without altering the x30 value. Directly pushing the return address onto the stack would require replacing an additional instruction, which we cannot afford. Our solution is to design a dedicated trampoline for each SVC that pushes the address of the instruction following the SVC onto the stack upon entering the trampoline. This way, after executing the SVC in the trampoline, we can retrieve the return address from the stack. Additionally, we have minimized the code length required for each SVC instruction’s individual trampoline. Given that the number of SVC instructions in the process image of typical applications is relatively small, our trampolines do not introduce significant memory and virtual memory overhead.

The second issue is that in the first replacement scheme, the first instruction is replaced with an assignment to x8, which has a small immediate value range, limiting the jumpable virtual address range to 0 to 65535, accommodating only 16383 instructions. To efficiently use these virtual addresses, we designed a multilevel trampoline jump scheme.

As shown in Figure \ref{fig:figure2}, Our trampoline design consists of three levels. The first and second levels are dedicated trampolines for each SVC instruction, and the final level is a shared trampoline for all system calls.

The first-level trampoline addresses the issue of the small virtual address range. For the first jump strategy, the jump address range is from 0 to 65535. These virtual addresses are very valuable, so the main purpose of this level of trampoline is to exit as quickly as possible to minimize virtual address usage. We use three mov instructions and one br instruction to form the first-level trampoline. These three mov instructions load a 48-bit immediate value into x8, which is the address of the second-level trampoline. The 48-bit immediate value allows the second-level trampoline to be placed anywhere in the virtual address space.

Each SVC instruction has its own second-level trampoline to handle operations related to each SVC. In this trampoline, the previously replaced assignment instruction is re-executed first to restore the original value of the x8 register. Next, the return address of the system call, which is the address of the instruction following the SVC, is pushed onto the stack. This allows the shared third-level trampoline to retrieve the return address from the stack. Finally, control is transferred to the third-level trampoline.

The third-level trampoline is shared by all monitored objects. Upon entering the trampoline, the register context is first saved. Next, the user-provided hook function is called. After that, the register state is restored, and the original system call is executed. Finally, the return address is retrieved from the stack, and control is returned to the original instruction flow.

\subsection{Completeness Strategy}
\label{sec:section3.3}
We have designed three Completeness strategies to ensure the  security and thoroughness of our tool. As shown in Figure \ref{fig:combined}(c), these strategies aim to handle the issue where the jump target address appears between the two replaced instructions (Including the original SVC instruction, which was replaced by the BR instruction in Figure \ref{fig:combined}(c). If this occurs, it will result in the failure to execute the first replacement instruction that assigns the address of the first-level trampoline to x8, only executing the br x8 instruction and causing a jump to an incorrect address.
\begin{figure}[!t]
  \centering
  \includegraphics[width=0.5\textwidth]{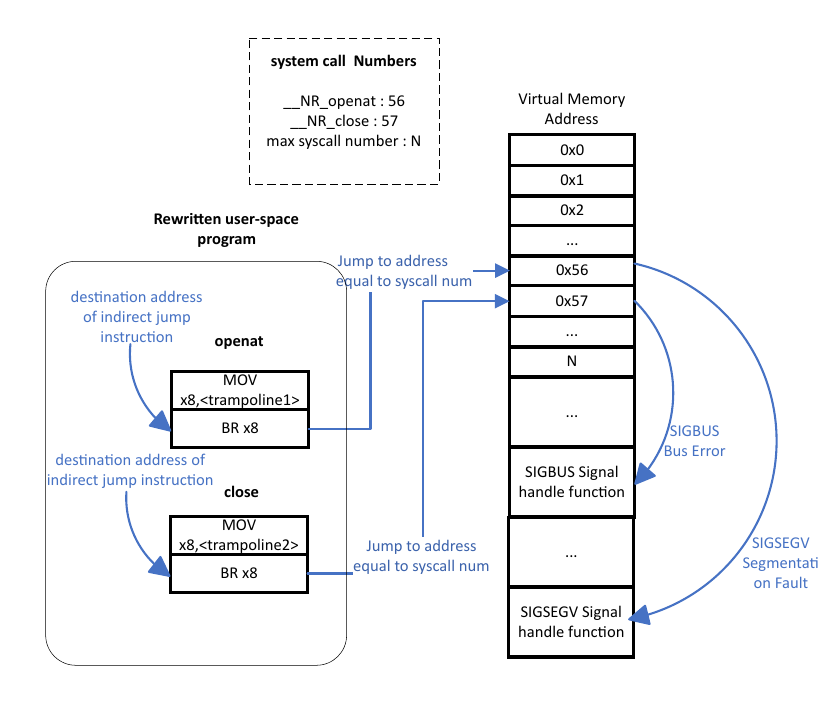} 
  \caption{Execution flow when the target address of an indirect jump is an SVC instruction.}
  \Description{A brief description of the image.}
  \label{fig:figure3}
\end{figure}
Our strategy is to replace such SVC instructions with brk or illegal instructions. When these instructions are executed, they generate a signal that we intercept using a registered signal handler. We have designed three completeness strategies to identify the SVC instructions that need to be replaced with either brk or an illegal instruction.

\textbf{The first completeness strategy.} We inspect a portion of the instructions preceding each system call SVC instruction. If there is no assignment instruction to the x8 register, if there are clear ABI omissions, or if there are jump instructions between the two replaced instructions, we consider these missing ABIs are likely executed in another function. Therefore, we intercept these SVC instructions using signals.

\textbf{The second completeness strategy.} For direct jump instructions (such as the b or bl instruction), as shown in Figure \ref{fig:combined}(a), we inspect all direct jump instructions in the process's memory and calculate the target address by adding the PC and the offset. If these jump addresses fall between the two instructions, we change the system call interception method to use signals.

For indirect jumps (i.e., in the ARM architecture, the instructions that perform jumps using registers), we provide an interface that allows users to specify locations where they believe the SVC instruction should be replaced with brk or illegal instructions through a configuration file. This approach is feasible because most system calls are based on open-source libraries (e.g., glibc), and users may have information about the targets of indirect jumps (such as function pointer destinations). Users can provide information based on system call numbers, the library's path in the file system and instruction offsets within the library, or virtual addresses. We compare this information with the detected SVC instructions and, if they match, replace the SVC with brk or illegal instructions.

\textbf{The third completeness strategy.} As shown in Figure \ref{fig:figure3}, since we only replace the SVC instruction and the instruction that assigns a value to the x8 register, we only need to consider the scenario where the indirect jump target address falls between these two instructions. This means the first instruction that assigns the address of the first-level trampoline to x8 was not executed, leaving the x8 register holding the system call number, which is less than 600. Additionally, the address from 0 to 4095 is left empty, because the first-level trampoline’s virtual address starts at 4096. Because the SVC instruction is replaced with the br x8 instruction, executing this instruction will result in a bus error or segmentation fault. We intercept signals caused by segmentation faults and bus errors. In the signal handler, because the PC value equals the x8 register value, which equals the system call number, we append the problematic system call number to the configuration file. Additionally, because most indirect jumps will use the BLR instruction, which saves the return address (the next instruction after BLR) in the x30 register, We use the value in x30 to locate the BLR instruction. Then, using the destination register of the BLR instruction, we locate the SVC instruction. We compare the SVC instruction's address with the load addresses of various libraries to determine the library and offset of the SVC instruction. Finally, we write these information to the configuration file. After completing this analysis, we re-execute the application. When the application is restarted the second time, it will automatically read the information from configuration file and avoid the previous error.

The above methods are effective because such situations are extremely rare, and the implementation of system calls is mostly concentrated in specific libraries (e.g., glibc). Once the above situation is identified, the information is generally applicable and shareable for most applications running on the same host or even over the network, as long as the library versions are the same. Essentially, this configuration only needs to be done once, and as long as the libraries are not upgraded, similar situations will not occur again.

We use configuration files to control the behavior of our Completeness strategies. Due to the system call ABI, it is unusual for a jump target address to fall between the assignment to the x8 register and the SVC instruction. The primary purpose of our Completeness policy is for insurance. So these Completeness strategies are disabled by default. By changing the configuration file, these Completeness strategies can be enabled separately, specifying the SVC instruction locations to be intercepted by signals and whether to use illegal instructions or brk instructions to generate signals.

\subsection{Implementation Details}
We use LD\_PRELOAD to ensure our library executes before the application. First, we use procfs to view the entire address space of the process and perform a linear scan of the code sections using the libopcodes library from GNU Binutils, recording all SVC instruction information and replacing them according to our replacement strategy. Next, we use mmap to allocate virtual memory for the first two levels of trampolines and fill the allocated virtual memory with predefined binary instruction codes. After filling, we use mprotect to protect them.

It is worth noting that we chose 4096 as the starting address for the first-level trampoline, rather than 0. Typically, memory access to virtual address 0 (i.e., the NULL pointer) triggers a page fault due to the lack of physical memory mapping, leading to the termination of the user-space program. This is crucial for preventing faulty programs from continuing to run. If we were to choose 0 as the virtual address, it would prevent NULL pointer access errors in user-space programs. Therefore, we leave the 0 address vacant, and since mmap addresses must be 4KB aligned, 4096 is chosen. 

Additionally, when a segmentation fault or bus error is triggered because the SVC instruction was replaced with br x8 and the indirect jump target address falls between the two replaced instructions, the PC value equals the x8 register value, which is the system call number. In the signal handler for segmentation faults and bus errors, we leverage this characteristic to distinguish such cases from memory access to virtual address 0, as well as other segmentation faults or bus errors caused by the program's own logic. Specifically, we check whether the PC value is less than the maximum system call number and whether it matches the value in the x8 register.

The third-level trampoline is an assembly-written trampoline function. Its execution flow includes: saving the register context, executing the user-provided hook function, restoring the register context, executing the system call, and returning. It is important to note that when the user-provided hook function calls the function that encapsulates the SVC instruction, it would fall into an infinite loop because the SVC instruction has been replaced, and the replaced code would bring the execution back to the hook function.

To avoid this issue, we use dlmopen, which can load the library into a separate namespace, thus preventing the user-provided hook function from associating with the library already rewritten by ASC-Hook. We assume that ASC-Hook users will build the core implementation of the hook as a separate shared library. We use dlmopen to load this library and dlsym to obtain the pointer to the core implementation of the hook function, ensuring that executing the user-provided hook function does not result in an infinite loop.

\section{Evaluation}
\label{sec:section4.0}
In this study, we conducted an evaluation on the ARM platform, utilizing a experimental environment equipped with a dual-core Neoverse-N1 processor architecture clocked at 2.8GHz, 4GB of DRAM, running Linux with kernel version 5.4.0-174-generic and glibc version 2.31. We compared ASC-Hook with  ptrace, signal interception methods, and LD\_PRE
LOAD. The ptrace method uses kernel interfaces to intercept system calls; signal interception methods include using the brk instruction or illegal instruction replacements for the SVC instruction, which trigger SIGSEGV and SIGILL signals respectively, allowing system call interception by capturing these signals. This method is widely applied in our Completeness strategy, and in this experiment, it was used to intercept all of the system calls. System Call User Dispatch (SUD)\cite{bertazi2021syscall} was introduced in Linux 5.11, providing a way to redirect system calls to arbitrary user-space code. When SUD is activated, hook points send a SIGSYS signal to the user-space process, which essentially intercepts system calls through signals. However, since this mechanism is not implemented on the ARM architecture, we did not test it. The efficiency of these methods is generally consistent, with signal interception methods representing them in our experiments. LD\_PRELOAD represents function interception methods, characterized by high efficiency but limited by its inability to fully hook system calls.
\begin{table}[h!]
\centering
\begin{tabular}{|l|c|}
\hline
\textbf{Mechanism} & \textbf{Time [ns]} \\
\hline
LD\_PRELOAD & 6.79344 \\
\hline
Signal interception methods & 986.7024 \\
\hline
Ptrace & 2059.5956 \\
\hline
ASC-Hook & 33.52524 \\
\hline
\end{tabular}
\caption{The overhead for hooking a system call.}
 \label{tab:table2}
\end{table}

We used the getpid system call to measure the overhead of system call interception methods. To eliminate the kernel-crossing system call overhead, we used a hook function that returns a virtual value instead of executing the getpid system call. As shown in Table \ref{tab:table2}, LD\_PRELOAD has the lowest overhead, with an average interception overhead of 6.79 nanoseconds per call. The overhead of ASC-Hook is approximately five times that of LD\_PRELOAD, while the overhead of signal interception and ptrace are 29 times and 61 times that of ASC-Hook, respectively.

\begin{figure}[!t]
  \centering
  \includegraphics[width=0.50\textwidth]{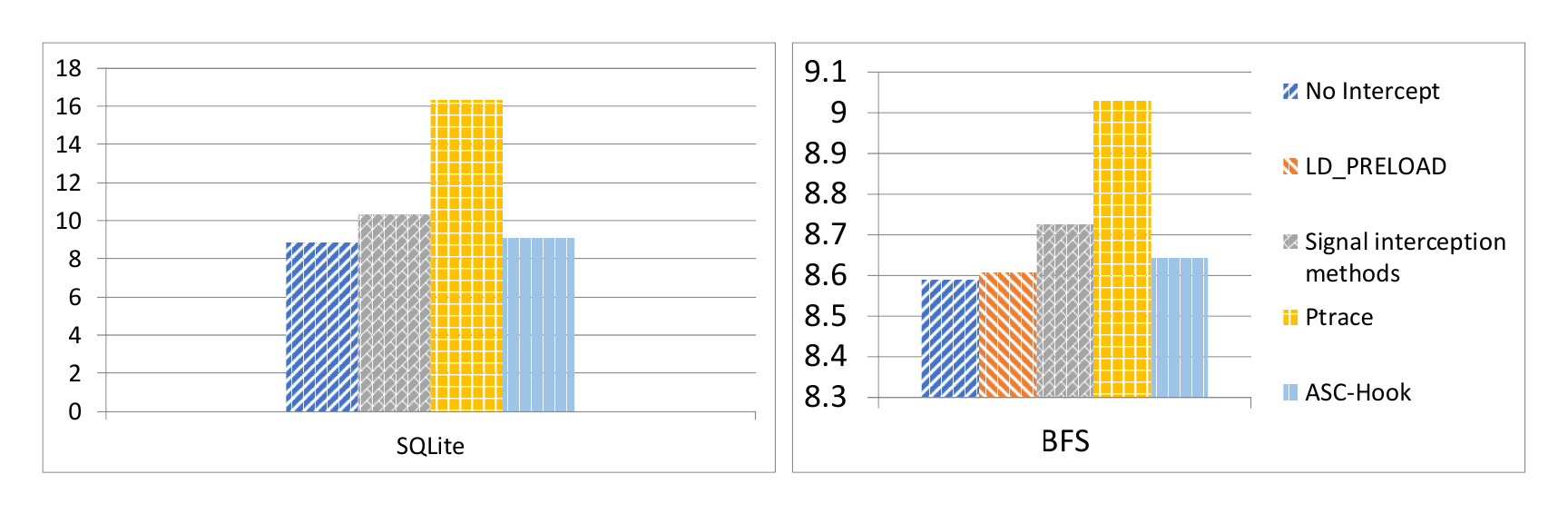}
  \caption{ Performance Impact of Interception Tools on SQLite and bfs}
  \Description{A brief description of the image.}
  \label{fig:figure8}
\end{figure}

To evaluate the impact of ASC-Hook on MPI parallel computing programs, we designed an MPI parallel computing program based on the BFS algorithm. The program reads a node matrix from a file and traverses it, involving numerous read operations that generate a large number of read system calls. Extensive testing of the MPI-based BFS benchmark program showed that the overhead of ASC-Hook accounts for 0.6\% of the total runtime of BFS. Additionally, we also tested interception methods, including LD\_PRELOAD, Signal interception method, and ptrace, with the experimental results shown in Figure \ref{fig:figure8}.

To systematically test the performance impact of ASC-Hook on various database operations, we chose the SQLite\cite{sqlite}  database. SQLite is an open-source, standalone, and highly reliable SQL database engine written in C, providing a fully-featured database solution. It can run on almost all mobile phones and computers and is embedded in numerous applications. We used SQLite's built-in benchmarking program, speedtest1, for testing. speedtest1 measures the speed of various common database operations, and we recorded the total execution time. The results are shown in Figure \ref{fig:figure8}. ASC-Hook introduced only 0.3 seconds of interception overhead, accounting for 3.3\% of the total runtime of speedtest1.

\begin{figure}[!t]
  \centering
  \includegraphics[width=0.50\textwidth]{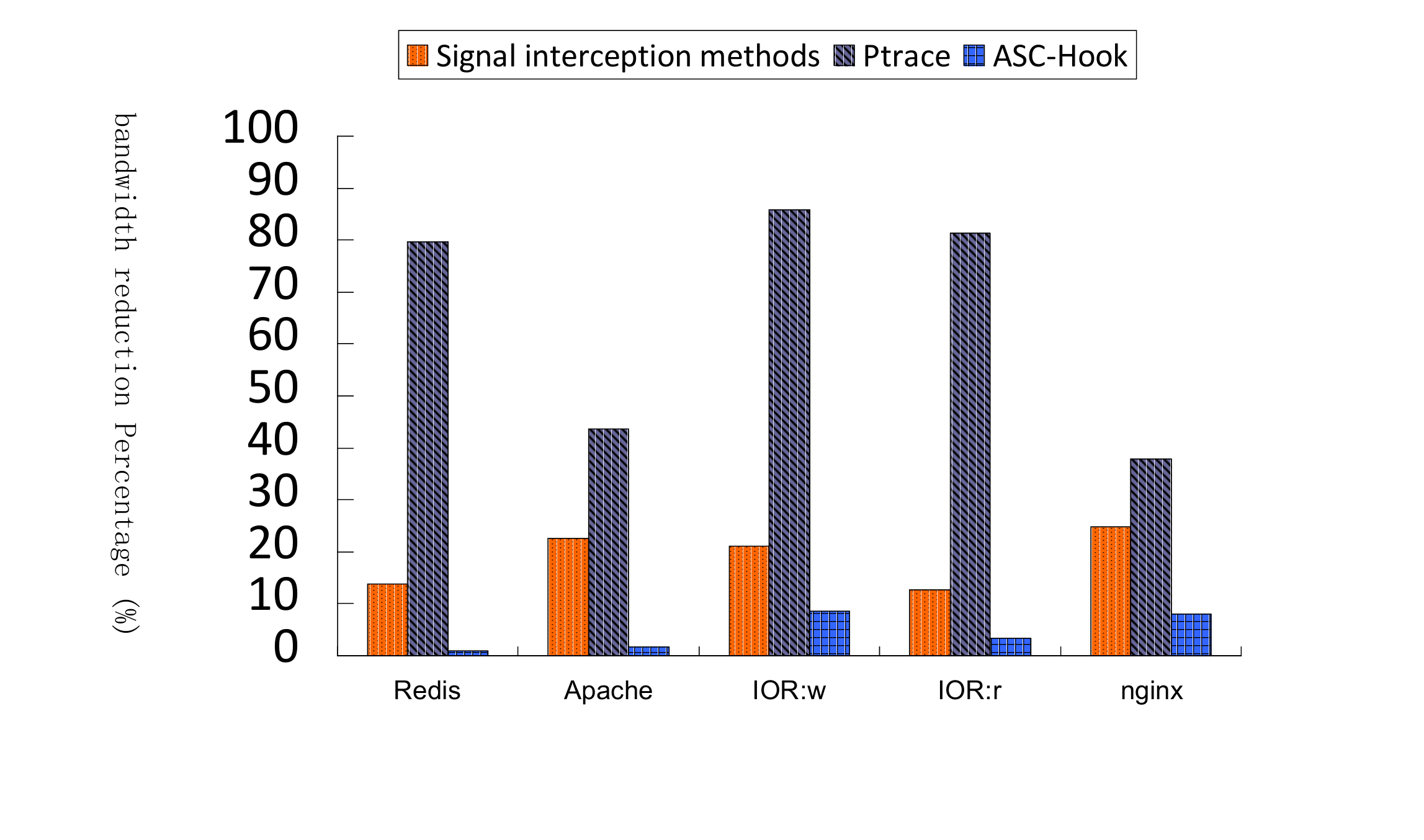}
  \caption{The impact of interception tools on bandwidth}
  \Description{A brief description of the image.}
 \label{fig:figure5}
\end{figure}
We tested the impact of system call interception tool on the bandwidth of real-world applications, measuring the performance of various tools by the rate of the bandwidth decrease percentage. The calculation formula is as follows:
\[
\text{Bandwidth Decrease Percentage} = \left( \frac{B_{\text{normal}} - B_{\text{tool}}}{B_{\text{normal}}} \right) \times 100\%
\]

where \( B_{\text{normal}} \) represents the bandwidth of the application running normally and \( B_{\text{tool}} \) represents the bandwidth with the tool applied.

We selected Redis\cite{sanfilippo2009redis}, Apache HTTP Serve\cite{apache}, InterleavedOrRandom (IOR)\cite{llnl2015ior} \cite{ior}, and Nginx\cite{nginx}  as our benchmark programs to evaluate the impact of signal interception, ptrace, and ASC-Hook on throughput. The results are shown in Figure  \ref{fig:figure5}.

\begin{itemize}
\item  Redis is an open-source, in-memory database that also serves as a non-relational database. It is widely used in scenarios requiring fast data access, such as caching, real-time analytics, and message queues. We chose Redis to evaluate ASC-Hook in real-world applications. To minimize the impact of inter-host communication on the overhead assessment, we conducted loopback tests on the same server, running the Redis server and executing redis\_benchmark, selecting the GET operation as the performance metric. 
\item 
Apache HTTP Server, commonly known as Apache, is one of the most popular open-source cross-platform web servers in the world. We used wrk for stress testing, sending continuous requests to the same static resource for 10 seconds with one thread and one connection. To minimize the impact of transmission distance on the network, we set up both the client and server on the same machine.
\item   
Interleaved or Random (IOR), developed by Lawrence Livermore National Laboratory (LLNL), is one of the most highly configurable and widely used I/O benchmarks. It is also an MPI application. We tested the impact of various tools on read and write bandwidth by writing 1 GB of contiguous bytes per task with a transfer size of 1 KB. 
\item   
Nginx is a high-performance HTTP and reverse proxy web server. We used the wrk client to send continuous requests to the same static resource for 10 seconds with one thread and one connection. The client and server were set up on the same machine.
\end{itemize}
For the applications mentioned above, in the order presented in Figure \ref{fig:figure5}, the bandwidth reduction Percentage caused by ASC-Hook are 0.96\%, 1.77\%, 8.52\%, 3.26\%, and 8\%, which are significantly lower than those caused by the other two methods.

\section{Limitations}

Firstly, due to our first replacement strategy requiring lower virtual addresses, ASC-Hook faces limitations when running on Windows and macOS. On Windows 10, VirtualAlloc fails when the specified virtual address is below 0x10000, making it impossible to apply ASC-Hook on Windows. On macOS, the virtual address 0 in user space is occupied by a special segment called \_\_PAGEZERO, preventing the application of ASC-Hook on macOS as well. However, if we completely avoid using the first replacement strategy and instead rely entirely on the second replacement strategy, it might solve this issue but would result in significant memory consumption.

Our tool uses LD\_PRELOAD to scan the process address space before the application executes and replaces SVC instructions. However, if a dynamically loaded library containing SVC instructions is loaded after this setup (e.g., if a library is opened with dlopen in the main function after the setup and one of its functions is called, which contains an SVC instruction.), it will not be hooked. We can address this issue by adopting the online binary rewriting approach proposed in X-Containers \cite{shen2019x}. This method captures system calls and rewrites the SVC instructions that trigger system calls in real time.

Since our tool registers signal handlers for illegal instructions, debug instructions, segmentation faults, and bus errors before executing the application, the application's own signal handlers can overwrite those registered. However, as mentioned earlier, even without using signals for interception, our first two replacement strategies can meet the needs of most applications. Moreover, most applications at least do not intercept signals generated by debug instructions.

The Virtual Dynamic Shared Object (vDSO) enables user space programs to access certain system calls directly by exposing their implementation through the kernel. Similar to other system call interception mechanisms, ASC-Hook does not support hooking vDSO-based system calls. However, we can disable vDSO to address this issue. 

In edge cases, ASC-Hook may encounter some performance bottlenecks. For example, if the total number of SVC instructions for a process exceeds 3,840, or if brk or illegal instructions are repeatedly executed, the overhead could become significant. However, as shown in Tables \ref{tab:table1} and \ref{tab:svc_instructions_compact} and the evaluation section, these are extreme cases and rarely occur in practice.

\section{Related Work}

For disassembly, in addition to the linear scanning method used in this paper, another common static approach is recursive traversal. This method starts with a linear sweep and switches to either a depth-first or breadth-first search upon encountering a branch instruction, continuing the disassembly from there. Hybrid approaches combining multiple strategies also exist\cite{nanda2006bird}. Notably, ELISA introduces a supervised learning-based disassembly method supporting ARM architecture, efficiently identifying code segment boundaries and separating inlined data blocks with high precision\cite{de2018elisa}. These methods have the potential to be integrated into our tool to enhance its handling of inlined data.

Some works intercept system call wrapper functions rather than directly intercepting system calls\cite{lei2023birfia} \cite{ldpreload2024}. This type of solution has minimal performance impact but lacks thoroughness because system call instructions may appear outside of wrapper functions, or multiple system call instructions may be invoked within a single function. Additionally, function-level interceptors must identify all system call wrapper functions and map them to the actual system calls executed, which is challenging to scale in practice.

In the kernel, existing mechanisms such as the PTRACE API on Linux\cite{padala2002ptrace} and the application debugging API on Windows \cite{kath1992debugging} can monitor each system call and retrieve related register information.However, the frequent switching between user mode and kernel mode, as well as the context switching between the target process and the debugger, results in significant overhead. The Berkeley Packet Filter (BPF \cite{mccanne1993bsd}) intercepts system calls efficiently and comprehensively by inserting user code at preset key points in kernel space. However, BPF programs are not capable of modifying or simulating the behavior of system calls.

In the field of dynamic binary rewriting, there are several well-known tools, including Dyninst (dynamic version) \cite{bernat2011anywhere}, DynamoRIO \cite{bruening2004efficient}, and Pin \cite{10.1145/1065010.1065034}. Dyninst uses Linux's ptrace tool to perform dynamic binary instrumentation on running processes, but it introduces significant runtime and memory overhead. DynamoRIO and Pin intercept application execution by copying and transferring instructions from the original instruction set to a code cache that can be safely modified. This process allows developers to insert analysis, monitoring, or optimization code without altering the original program. However, the processes of runtime instruction translation, parsing, cache management, and execution path redirection result in significant time overhead, while maintaining an additional code cache incurs memory overhead.

zpoline\cite{288689} replaces the two-byte syscall/sysenter instructions with two-byte callq rax instructions, achieving efficient system call interception. However, due to the requirement for instruction alignment and some other reasons, this method cannot be applied to the ARM architecture.

\section{Conclusion}
This paper introduces ASC-Hook, a system call hooking mechanism for ARM architecture. It ensures efficient and comprehensive interception through a novel hybrid design. We extensively tested ASC-Hook with real-world applications, demonstrating its comprehensive and efficient interception capabilities. The experimental results indicate that ASC-Hook maintains an average performance overhead of 3.7\% for most system call-intensive applications.

% \bibliographystyle{ACM-Reference-Format}
%\bibliography{sample-basemy}

\appendix

\end{document}